\documentstyle[12pt]{article}
\begin{document}
\newcommand{\siml}{\stackrel{<}{\sim}}
\newcommand{\simg}{\stackrel{>}{\sim}}

 \newcommand \be {\begin{equation}}
\newcommand \ee {\end{equation}}
 \newcommand \ba {\begin{eqnarray}}
\newcommand \ea {\end{eqnarray}}
\def\oppropto{\mathop{\propto}}
\def\operarrow{\mathop{\longrightarrow}}
\def\opsimeq{\mathop{\simeq}}

\title{\bf PEIERLS INSTABILITY OF VORTEX TUBES.}

\vskip 3 true cm

\author{Jean-Philippe Bouchaud$^1$}

\date{\it
$^1$ Service de Physique de l'\'Etat Condens\'e,
 Centre d'\'etudes de Saclay, \\ Orme des Merisiers, 
91191 Gif-sur-Yvette C\'edex, FRANCE \\ 
e-mail: bouchaud@amoco.saclay.cea.fr}

\vskip 1 true cm

%\date{\today}

\maketitle

\begin{abstract}
We discuss the possibility of a new low temperature instability of vortex tubes
to a `folded' state, driven by the coupling to the normal electron states inside the cores. The basic mechanism is that a bended tube creates an effective potential for the electrons, which destabilizes the tube in a way reminiscent of the usual Peierls instability. 

R\'esum\'e: Nous discutons la possibilit\'e d'une nouvelle instabilit\'e `ondulatoire' des tubes de vortex, induite par le couplage aux \'etats \'electroniques normaux au coeur de ceux-ci. Le m\'ecanisme responsable de cette instabilit\'e est le fait qu'un tube ondul\'e cr\'ee sur les \'electrons qui y sont confin\'es un potentiel effectif attractif. L'instabilit\'e qui en r\'esulte s'apparente \`a l'instabilit\'e de Peierls.

\end{abstract}

\vfill

\vskip 2 true cm

\noindent  \hfill March 1996

\newpage

{\centerline{\bf SUR UNE INSTABILITE DE PEIERLS DE VORTEX}}
\vskip 1cm

{\centerline{ VERSION FRANCAISE ABREGEE}}

\vskip 1cm

La physique des lignes de vortex dans les supraconducteurs a attir\'e une attention consid\'erable ces derni\`eres ann\'ees. Les propri\'et\'es {\it collectives} de ces objets, en particulier en pr\'esence de d\'esordre, ont
\'et\'e explor\'ees en grand d\'etail (Blatter et al., Fisher et al.). Simultan\'ement, de nouvelles 
techniques exp\'erimentales ont \'et\'e d\'evelopp\'ees pour sonder la structure
microscopique de ces vortex (Maggio-Aprile et al.).  

Le but de cette note est d'explorer les cons\'equences d'un effet,
discut\'e r\'ecemment par Goldstone et Jaffe, qui est le suivant: un \'electron confin\'e dans un tube courb\'e ressent un potentiel effectif
n\'egatif proportionel au carr\'e de la courbure locale (Eq.(1)). Ceci pourrait avoir comme
cons\'equence interessante de rendre les vortex dans les supraconducteurs
\`a faible longueur de coh\'erence instables vis \`a vis de petites
modulations p\'eriodiques. Le m\'ecanisme de cette instabilit\'e est le
couplage entre g\'eom\'etrie du vortex et degr\'es de libert\'e
longitudinaux des \'electrons normaux confin\'es dans le coeur du vortex (Caroli et al.).
Une distortion p\'eriodique du vortex cr\'ee donc un potentiel de
p\'eriode double, qui abaisse consid\'erablement l'\'energie du gaz
electronique unidimensionel (Eq.(3)). En revanche, cette distortion coute une
\'energie `\'elastique' due \`a l'allongement du vortex et \`a la
modification du champ magn\'etique environnant (Eq.(4)). De fa\c con tout \`a fait
similaire \`a l'instabilit\'e de Peierls, cette comp\'etition conduit \`a
l'apparition d'une modulation de vecteur d'onde $k_f$, et d'amplitude
$\delta$ dont l'ordre de grandeur vaut, par rapport \`a la longueur de
coh\'erence $\xi$, $\frac{\delta}{\xi} \simeq (k_f \xi)^{-2}$. Cette
instabilit\'e ne conduit \`a des effets appr\'eciables que si $k_f \xi
\siml 1$, ce qui assure que le gaz electronique est bien unidimensionel
et que le couplage avec la conformation externe du vortex est important.
De plus, la temp\'erature doit \^etre tr\`es faible, pour \'eviter que
les fluctuations thermiques (divergentes avec la longueur du vortex -- voir Eq. (6))) ne
masque la modulation p\'eriodique discut\'ee ici. Pour des vortex en interaction, la longueur effective des vortex est fix\'ee par le module de cisaillement du reseau d'Abrikosov, et est de l'ordre du micron. Pour $T=10^{-2}$ K, les fluctuations thermiques sont de l'ordre de quelques Angstroems, d\'eja comparables \`a $(k_f \xi)^{-2}$. Une instabilit\'e analogue
pourrait aussi apparaitre pour d'autres objets unidimensionnels, comme les dislocations.

\vskip 2cm

The physics of vortex lines in superconductors is fascinating. Many new
aspects have been discussed in the recent years in connection with the high
temperature materials, and various phases discovered (vortex liquid, vortex
glass, etc. (Blatter et al., Fisher et al.), unveiling interesting collective properties of these lines
defects. Simultaneously, new experimental techniques (tunneling microscopy) have been developed to
investigate the microscopic structure of the vortex cores (Maggio-Aprile et al.) , and to probe the
normal electron states within these cores (Caroli et al.). 

The aim of this note is to show that individual vortex lines should, under certain conditions, undergo a Peierls like transition to a `folded' state, 
driven by the coupling between these normal electron states within the core
and the conformation of the core. The idea is very simple and relies on the
fact that the effective Schrodinger equation describing an electron confined
in a distorted tube contains a curvature induced negative potential term. As
shown by Goldstone and Jaffe, in the limit of small, slowly varying
curvature, the Schrodinger equation reads:
$$
- \frac{\hbar^2}{2m} [\nabla^2_\perp \phi] - \frac{\hbar^2}{2m}
\frac{\partial^2 \phi}{\partial s^2} - \frac{\hbar^2}{8m} [\kappa(s)]^2 \phi
= E \phi \eqno(1)
$$
where $s$ is the curvilinear coordinate along the distorted tube,
$\kappa(s)$ the local curvature of the tube, and $\nabla_\perp$ is the
Laplacian in the directions transverse to the tube. Writing $\phi(s,\vec
x_\perp)=\psi(\vec x_\perp) \varphi(s)$, one finds a one
dimensional Schrodinger equation for $\varphi(s)$, with an effective
attractive potential ${\cal V}(s)=-\frac{\hbar^2}{8m} [\kappa(s)]^2$. As
emphasized by  Goldstone and Jaffe, bends are thus sufficient to create bound
states in a tube. 

Now consider the one dimensional gas of normal electrons living in a vortex
tube. If the tube is periodically modulated in a given direction, say $\vec
x_\perp(s)=(\delta \cos(Qs),0)$, the effective potential is given by:
$$
{\cal V}(s)=-\frac{\hbar^2}{16 m} \delta^2 Q^4 [1+\cos(2Qs)] \eqno(2)
$$
Note that the curvature $\delta Q^2$ cannot exceed $\xi^{-1}$, where 
$\xi$ the size of the vortex core. Furthermore, Goldstone and Jaffe's description only hold if the transverse boundary conditions confining the electron are sufficiently `sharp'. Since the `width' of the vortex walls is also of order $\xi$, this imposes that $\delta \simg \xi$. This condition will restrict our results to the case of {\it short coherence length} superconductors, such that $k_f \xi$ is not too large, where $k_f$ is the Fermi momentum.

The change of energy (per unit length) of
the electron gas to this periodic potential is given by the classical
expression (Kagoshima et al.): 
$$
\Delta E_{1} = - \rho \frac{\hbar^2}{16 m} \delta^2 Q^4
- \rho \chi(2Q) [\frac{\hbar^2}{16 m} \delta^2 Q^4]^2 \eqno(3)
$$
where $\rho$ is the linear density of electrons within the core. In the case  where $k_f \xi \siml 1$, $\chi(Q)$ the one dimensional polarisation function, which diverges at $Q=2k_f$. Eq. (3) means that the effective free energy of the line will contain {\it negative} terms proportional to the curvature and the square of the curvature.

On the other hand, the vortices have a {\it positive} line tension $c_{44}$, reflecting the fact that the superconducting order parameter is modified around the vortex. For a single vortex line, one finds $c_{44} \simeq (\frac{\phi_0}{4 \pi
\lambda})^2$, with $\phi_0$ the flux quantum, $\lambda$
the penetration depth. The energy per
unit length of the distorsion then reads:
$$
\Delta E_{2} = + \frac{1}{4} c_{44}\delta^2 Q^2 \eqno(4)
$$
$c_{44}$ actually depends on the wavevector of the distorsion (Brandt et al.), and on the fact that other vortices are present. In particular, one also expects a positive curvature term of the order of $c_{44}\delta^2 Q^2 (Q\xi)^2$ to appear in Eq. (4).

Let us first look at long wavelength modulations $Q \ll k_f,\xi^{-1}$, such that
$\chi(2Q) \simeq 0$. Comparison between Eqs. (3) and (4) reveals that modes
with $Q>Q^*$ are unstable, with 
$$
\rho \frac{\hbar^2}{2 m}Q^{*2} =  2 c_{44}(Q^*) \eqno(5)
$$
Taking $\xi = 1 nm$, an electron density $\rho = 1 \ nm^{-1}$, a Fermi
energy $\frac{\hbar^2}{2 m} k_f^2 = 100 K$, and $c_{44} \simeq 50 K nm^{-1}$ (corresponding to $\lambda = 1 \mu m$), one finds that $Q^*$ is $ \simeq k_f$, which means that the instability will actually be of short
wavelength, and $\chi(2Q)$ cannot be neglected. The instability thus takes
place where $\chi(2Q)$ is maximum, i.e. right at $Q=k_f$ (and not $2k_f$ as in
the usual Peierls instability). The resulting value of the modulation $\delta$
requires higher order terms in Eqs.(3) and (4), but presumably is of order
$(k_f^2 \xi)^{-1}$. Correspondingly, a gap in the longitudinal electron states for
$k=k_f$ will appear. Note that $\frac{\delta}{\xi} \simeq (k_f \xi)^{-2}$, which in turn imposes the condition $k_f \xi \siml 1$ mentioned above. For larger $\xi$, the electron gas inside the core loses its one-dimensional nature, and the influence of the modulated boundary condition becomes quickly irrelevant.

The above discussion was restricted to zero temperature. When the
temperature grows, $\chi(2k_f)$ becomes finite and this reduces the
tendency to instability. Nevertheless, the first term in $\Delta E_1$ still makes
the high $Q$ modes unstable, due to the basic effect that the curvature
pushes down all the electronic states. However, the effect discussed here
should rapidly become unobservable because of the thermal fluctuations of the
vortex. For an isolated vortex, the thermal displacement is given by (Blatter et al., Fisher et al.):
$$
\delta_{T} \propto  \sqrt{\frac{{\cal L} T}{c_{44}}}\eqno(6)
$$
where $\cal L$ is the length of the vortex. For interacting vortices, $\cal
L$ is replaced by a length determined by the shear modulus of the vortex
lattice, and is of the order of $1 \mu m$. For $T=10^{-2}$ K, one thus
finds $\delta_{T}$ of the order of a few Angtroems,
already comparable to $(k_f^2 \xi)^{-1}$. Hence, the effect suggested here
might only be observable at very low temperatures. A similar instability might also be relevant for other linear defects, such as dislocations.

\vskip 2cm
Acknowledgments. This work was motivated by a discussion with M.
Feigel'man. I also thank A. Comtet for interesting remarks, and for
providing the paper by Goldstone and Jaffe, and P.G. de Gennes for critical comments.
\vskip 2cm
REFERENCES:
\vskip 1cm

For an extensive review, see G. Blatter, M. V. Feigel'man, V. B. Geshkenbin,
A. I. Larkin and V.M. Vinokur, 1994, `Vortices in high temperature superconductors', Rev. Mod. Phys. {\bf 66} 4.

See e.g. E.H. Brandt, U. Essmann, 1987, `The flux-line lattice in Type-II superconductors', Phys. Status Solidi {\bf B 144}, 13 

C. Caroli, P. G. de Gennes, J. Matricon, 1964, `Bound fermion states on a vortex line in Type-II superconductors', Phys. Letters {\bf 9}, 307 

D.S. Fisher, M.P.A. Fisher, D.A. Huse, 1991, `Thermal fluctuations, quenched disorder, phase transitions and transport in Type-II superconductors' Phys. Rev. {\bf B 43}, 130 

J. Goldstone, A. Jaffe, 1992, `Bound states in twisted tubes', Phys. Rev. {\bf B 45}, 14100 

See e.g. `One Dimensional Conductors', by S. Kagoshima, H. Nagasawa, T.
Sambongi, Solid-State Science, Springer-Verlag (1991).

I. Maggio-Aprile, Ch. Renner, A. Erb, E. Walter, O. Fischer, 1995, `Direct vortex lattice imaging and tunneling spectroscopy of flux lines in YBACuO', Phys. Rev. Letters, {\bf 75}, 2754

\end{document}